\documentclass[aps,prc,superscriptaddress,twocolumn]{revtex4}
\usepackage{latexsym}
\usepackage{amssymb}
\usepackage{amsmath}
\usepackage{graphicx}
\usepackage{bm}
\usepackage{epsf}
\usepackage{epstopdf}
\usepackage{color,xcolor}

\usepackage[bookmarks=false,unicode=false,dvipdfm]{hyperref}
\hypersetup{hidelinks,colorlinks=true,allcolors=black,pdfstartview=Fit,breaklinks=true}

\newcommand{\orcidJi}{\href{https://orcid.org/0000-0001-6044-252X}{{\hspace{-0.8mm}{\color{white}{1}}\hspace{-1.5mm}}{\includegraphics[scale=0.05]{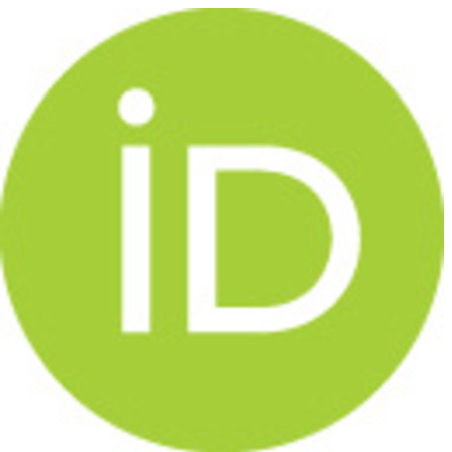}}}}
\newcommand{\orcidHu}{\href{https://orcid.org/0000-0002-1709-0159}{{\hspace{-0.8mm}{\color{white}{1}}\hspace{-1.5mm}}{\includegraphics[scale=0.05]{ORCIDiD.eps}}}}
\newcommand{\orcidShen}{\href{https://orcid.org/0000-0003-2717-9939}{{\hspace{-0.8mm}{\color{white}{1}}\hspace{-1.5mm}}{\includegraphics[scale=0.05]{ORCIDiD.eps}}}}

\begin{document}

\title{Nuclear pasta and symmetry energy in the relativistic point-coupling model}

\author{Fan Ji\hspace{1mm}\orcidJi}
\affiliation{School of Physics, Nankai University, Tianjin 300071, China}
\author{Jinniu Hu\hspace{1mm}\orcidHu}~\email{hujinniu@nankai.edu.cn}
\affiliation{School of Physics, Nankai University, Tianjin 300071, China}
\author{Hong Shen\hspace{1mm}\orcidShen}~\email{shennankai@gmail.com}
\affiliation{School of Physics, Nankai University, Tianjin 300071, China}

\begin{abstract}
Nonuniform structure of low-density nuclear matter, known as nuclear pasta, is expected
to appear not only in the inner crust of neutron stars but also in core-collapse supernova
explosions and neutron-star mergers. 
We perform fully three-dimensional calculations of inhomogeneous nuclear matter
and neutron-star matter in the low-density
region using the Thomas-Fermi approximation. The nuclear interaction is described in the
relativistic mean-field approach with the point-coupling interaction, where the meson
exchange in each channel is replaced by the contact interaction between nucleons.
We investigate the influence of nuclear symmetry energy and its density dependence
on pasta structures by introducing a coupling term between the isoscalar-vector and
isovector-vector interactions. It is found that the properties of pasta phases in
the neutron-rich matter are strongly dependent on the symmetry energy and its slope.
In addition to typical shapes like droplets, rods, slabs, tubes, and bubbles,
some intermediate pasta structures are also observed in cold stellar matter with a relatively large proton fraction.
We find that nonspherical shapes are unlikely to be formed
in neutron-star crusts, since the proton fraction obtained in $\beta$ equilibrium
is rather small. The inner crust properties may lead to a visible difference in
the neutron-star radius.
\end{abstract}

\maketitle

\section{Introduction}

A core-collapse supernova explosion is one of the most spectacular events in the universe,
which marks the violent death of a massive star. After the explosion, it leaves behind
either a neutron star or a black hole depending on its mass~\cite{Latt04,Burr06,Sumi06}.
Nuclear matter in supernovae and neutron stars covers a wide range of baryon density,
temperature, and isospin asymmetry~\cite{Oert17,Shen11}.
In the interior of neutron stars, the uniform matter is highly isospin asymmetric, while
its baryon density may be as high as several times nuclear saturation density $\rho_0$.
From the inside to the outside of a neutron star, the matter density decreases to subnuclear
region and the core-crust transition occurs when homogeneous matter becomes unstable against
the formation of nuclear clusters. It is believed that a neutron star consists of an inner crust of
nuclei in a gas of neutrons and electrons, as well as an outer crust of nuclei in a
gas of electrons without dripped neutrons~\cite{Cham08,Heis00,Webe05}.
The inner crust of neutron stars has attracted much attention due to its complex phase structure
and important role in astrophysical observations~\cite{Rave83,Stei08,Gril12,Bao15,Okam13,Fatt17}.
With increasing density in the inner crust, spherical nuclei become unstable
and the favored geometric shape may change from droplet to rod, slab, tube, and bubble
before the crust-core transition. These exotic nuclear shapes are known as pasta phases, which are expected to appear not only in the inner crust of neutron stars but also
in core-collapse supernova explosions and neutron-star mergers.
In warm stellar matter relevant for supernovae, light clusters like $\alpha$
particles and deuterons may be formed before the formation of heavy nuclei
and nuclear pasta~\cite{Oert17,Shen11,Pais15}.
It has been found that the presence of nuclear pasta in core-collapse supernovae
could alter the late-time neutrino signal and affect the evolution of a protoneutron
star~\cite{Rogg18,Horo16}.
The elastic properties of nuclear pasta are currently interesting to some researchers
for their relevance to the gravitational-wave searches from supernovae and neutron-star
mergers~\cite{Abbo17b,Peth20,Peth19,Capl18}.

During the last decades, the properties of pasta phases have been studied
by using various methods, such as the liquid-drop model~\cite{Rave83,Wata00,Bao14a,Pais15}
and the Thomas-Fermi approximation~\cite{Gril12,Bao15,Oyam93,Oyam07,Mene08,Mene10,Grill14}.
In these calculations, the Wigner-Seitz approximation was generally employed,
where typical geometric shapes of nuclear pasta were assumed in order
to simplify the calculations. However, the assumption of geometric symmetry would
artificially reduce the configuration space, and as a result, other possible pasta
shapes besides the typical structures may be missed. For a more realistic description
of pasta phases, there are some studies that have not explicitly assumed any geometric
shape and performed fully three-dimensional calculations for nuclear pasta based
on the Thomas-Fermi approximation~\cite{Will85,Okam12,Okam13}, Hartree-Fock
approach~\cite{Magi02,Newt09,Pais12,Schu13,Sage16,Fatt17}, and molecular dynamics
method~\cite{Capl17,Maru98,Sono08,Wata09,Schn13,Schn14}.
In these calculations, not only the typical pasta shapes assumed in the Wigner-Seitz
approximation were reproduced but also other complex structures such as the waffle
phase were reported~\cite{Fatt17,Schn14}.

The properties of pasta phases appearing in the inner crust of neutron stars
can be significantly influenced by the nuclear symmetry energy and its density
dependence~\cite{Gril12,Bao15,Oyam07}. In recent years, the symmetry energy
has received great interest due to its importance for understanding many phenomena
in nuclear physics and astrophysics~\cite{Oert17,LiBA08,Bald16}.
It has been shown that the properties of neutron stars, such as the radius and
the crust structure, are sensitive to the symmetry energy $E_{\rm sym}$
and its slope parameter $L$~\cite{Gril12,Bao15,Oyam07,Duco10,Mene11,Prov13,Ji19,Duco08,Pais16a,Pais16b}.
Great efforts have been devoted to constraining the values
of $E_{\rm sym}$ and $L$ based on astrophysical observations and terrestrial nuclear
experiments~\cite{Tews17,Hebe13,Latt14,Hage15,Roca15,Birk17,Dani17}.
In Ref.~\cite{Oert17}, a sufficient number of constraints on the symmetry
energy parameters have been summarized, and the most probable values
for the symmetry energy and its slope at saturation density
were found to be $E_{\rm sym}=31.7\pm 3.2$ MeV and $L=58.7\pm 28.1$ MeV, respectively,
with a much larger error for $L$ than that for $E_{\rm sym}$.

The first detection of gravitational waves from a binary neutron-star merger,
known as GW170817, provides valuable constraints on the tidal
deformability~\cite{Abbo17,Abbo18,Abbo19}, which also restricts the radii of
neutron stars~\cite{Tews18,De18,Fatt18,Mali18,Zhu18}.
More recently, the gravitational-wave events, GW190425~\cite{Abbo190425}
and GW190814~\cite{Abbo190814}, were reported by the LIGO and Virgo Collaborations,
which may give important information for the equation of state (EOS) of dense matter.
The recent observation by the Neutron Star Interior Composition Explorer (NICER)
provided a simultaneous measurement of the mass and radius for
PSR J0030+0451~\cite{Rile19,Mill19}. These exciting developments in astrophysical
observations suggest relatively small radii of neutron stars, which is likely to
favor a small value of the symmetry energy slope $L$.
It is well known that there is a strong correlation between the symmetry energy
slope and the neutron-star radius~\cite{Latt13,Ji19}. Furthermore, other properties of
neutron stars, such as the crust structure and the crust-core transition, are also
directly affected by the symmetry energy and its slope~\cite{Gril12,Oyam07,Bao15}.
In Ref.~\cite{Oyam07}, the density region of nonspherical nuclei was calculated
by using a parametrized Thomas-Fermi approximation, which was found to be sensitive
to the symmetry energy slope $L$. In our previous work~\cite{Bao15}, a self-consistent
Thomas-Fermi approximation was employed to study the pasta structures presented in
the inner crust of neutron stars and we found that only spherical nuclei can be
formed before the crust-core transition for $L\geq 80$ MeV, whereas nonspherical
pasta phases may appear for smaller values of $L$ (e.g., $L=40$ MeV).
A similar calculation using the coexisting phases method in the quark-meson coupling
model with $L=69$ and $90$ MeV was performed and showed that only droplets could
present in the inner crust of neutron stars~\cite{Gram17}.

It is interesting to clarify the correlation between the symmetry energy slope
and nuclear pasta structures. In the present work, we perform fully three-dimensional
calculations without any assumption about the pasta shapes.
We carry out the calculation in a cubic box with periodic boundary conditions.
The pasta phase presented in neutron-star crusts is highly isospin asymmetric matter
in $\beta$ equilibrium, whereas the one occurred in supernovae is less asymmetric
and out of $\beta$ equilibrium. Therefore, we investigate the pasta structures for
stellar matter with a fixed proton fraction and neutron-star matter in $\beta$ equilibrium. In this study, all calculations are carried out at zero temperature for simplicity.
It is known that nonuniform structure in supernova matter exists at low
temperatures ($T<15\,\rm{MeV}$)~\cite{Oert17,Shen11,Ji20}, but the three-dimensional
calculations at finite temperature require much more computational time.
For the nuclear interaction, we employ the relativistic mean-field model with
point-coupling force (RMF-PC), which has achieved great success in describing
various phenomena in nuclear physics over the past
decades~\cite{Niko92,Rusn97,Burv02,Wata03,Chen07,Niks08,Daou09,Zhao10,Meng14,Ring14,Ring15,Zhao18}.
In the RMF-PC approach, the finite-range interactions through meson exchange in
Walecka-type models are replaced by corresponding zero-range interactions (point-coupling)
together with derivative terms.
In the present work, we use the PC-PK1 parametrization proposed by Zhao {\it et al.}~\cite{Zhao10},
which was determined by fitting to observables of 60 selected spherical nuclei
and provides a good description of ground-state properties for the nuclei all over
the nuclear chart~\cite{Meng14,Ring15,Zhao18}.
In order to examine the effect of the symmetry energy slope $L$ on nuclear pasta structures,
we generate a set of models with different values of $L$ at saturation density based on
the PC-PK1 parametrization by introducing an additional coupling term
between the isoscalar-vector and isovector-vector interactions, which corresponds
to the $\omega$-$\rho$ coupling in the finite-range RMF models~\cite{Bao14b}.
It has been found that this term plays a crucial role in
determining the density dependence of the symmetry energy and affecting neutron-star
properties~\cite{Mene11,Prov13,IUFSU,Horo01,Horo03,Bao14a,Bao14b,Bao15}.
By adjusting two parameters simultaneously, a given slope $L$ at saturation density $\rho_0$
can be achieved and the symmetry energy $E_{\rm{sym}}$ at average nuclear density
$\rho_B=0.12\,\rm{fm}^{-3}$ is fixed to the value predicted by the original PC-PK1 parametrization.
We note that all models in the set have the same isoscalar
properties and a fixed symmetry energy at $\rho_B=0.12\,\rm{fm}^{-3}$,
but they have different symmetry energy slope $L$.
The difference of $L$ does not significantly affect the properties of stable nuclei
except the neutron-skin thickness. Therefore, the set of models can provide
very similar description of finite nuclei and symmetric nuclear matter.
It is helpful to use these models to study the impact of the symmetry energy slope $L$
on nuclear pasta structures at subnuclear densities.

This article is organized as follows. In Sec.~\ref{sec:2},
we briefly describe the RMF-PC model employed in the three-dimensional
calculations of nuclear pasta phases. In Sec.~\ref{sec:3}, we discuss the
model parameters used in this study. In Sec.~\ref{sec:4}, we present the
numerical results and discuss the effects of the symmetry energy and its slope
on pasta structures. Sec.~\ref{sec:5} is devoted to the conclusions.

\section{Formalism}
\label{sec:2}

The nonuniform matter at subnuclear densities is studied within the Thomas-Fermi
approximation. Generally, the Wigner-Seitz approximation with several typical geometric
shapes is employed in the Thomas-Fermi calculation of nuclear pasta phases,
where the assumed geometric symmetry can help simplify the calculation to a one-dimensional
problem, but it artificially reduces the configuration space.
In the present work, we prefer to perform fully three-dimensional calculations without
any assumption about the geometric symmetry. The calculation is carried out in
a large cubic cell where the periodic boundary condition is used.
The nucleons in the cell tend to form clusters in order to lower the free energy of the system.
The electrons are assumed to be uniformed in the cell for simplicity, since the electron
screening effect caused by its nonuniform distribution is relatively small
at subnuclear densities~\cite{Maru05}.

For the nuclear interaction, we employ the RMF-PC approach, where the finite-range interactions
through meson exchange are replaced by corresponding zero-range interactions
together with derivative terms. For a system of nucleons and electrons,
the Lagrangian density of the RMF-PC model reads
\begin{eqnarray}
 \nonumber
 {\cal L} &=& {\bar{\psi}}\left(i\gamma_\mu\partial^\mu-m_{N}\right)\psi
 + \bar\psi_{e}\left(i\gamma_\mu\partial^\mu-m_{e}\right)\psi_{e} \\
 \nonumber
 & & -\frac{1}{2}\alpha_S\left(\bar\psi\psi\right)\left(\bar\psi\psi\right)
     -\frac{1}{2}\alpha_V\left(\bar\psi\gamma_\mu\psi\right)\left(\bar\psi\gamma^\mu\psi\right) \\ \nonumber
 & & -\frac{1}{2}\alpha_{TV}\left(\bar\psi\vec{\tau}\gamma_\mu\psi\right)\left(\bar\psi\vec{\tau}\gamma^\mu\psi\right) -\frac{1}{3}\beta_S\left(\bar\psi\psi\right)^3 \\ \nonumber
 & & -\frac{1}{4}\gamma_S\left(\bar\psi\psi\right)^4
     -\frac{1}{4}\gamma_V\left[\left(\bar\psi\gamma_\mu\psi\right)\left(\bar\psi\gamma^\mu\psi\right)\right]^2 \\ \nonumber
 & & -\gamma_C\left[\left(\bar\psi\gamma_\mu\psi\right)\left(\bar\psi\gamma^\mu\psi\right)\right]
              \left[\left(\bar\psi\vec{\tau}\gamma_\mu\psi\right)\left(\bar\psi\vec{\tau}\gamma^\mu\psi\right)\right] \\ \nonumber
 & & -\frac{1}{2}\delta_{S}\partial_\nu\left(\bar\psi\psi\right)\partial^\nu\left(\bar\psi\psi\right)
 -\frac{1}{2}\delta_{V}\partial_\nu\left(\bar\psi\gamma_\mu\psi\right)
                           \partial^\nu\left(\bar\psi\gamma^\mu\psi\right) \\ \nonumber
 & & -\frac{1}{2}\delta_{TV}\partial_\nu\left(\bar\psi\vec\tau\gamma_\mu\psi\right)
                            \partial^\nu\left(\bar\psi\vec\tau\gamma^\mu\psi\right)
     -\frac{1}{4}F^{\mu\nu}F_{\mu\nu}  \\
 & & -e\frac{1-\tau_3}{2}\bar\psi\gamma_\mu\psi A^\mu
                                     +e\bar\psi_{e}\gamma_\mu\psi_{e}A^\mu,
\label{LAG}
\end{eqnarray}
where $m_{N}$ and $m_{e}$ are the nucleon and electron masses, respectively.
$A^\mu$ is the four-vector potential of the electromagnetic field with
$F^{\mu\nu}$ being its antisymmetric field tensor.
The interactions include four-fermion terms with coupling constants $\alpha$,
while $\beta$ and $\gamma$ refer to third- and fourth-order terms, respectively.
The interactions with $\delta$ contain the derivative couplings.
The subscripts $S$, $V$, and $TV$ refer to isoscalar-scalar, isoscalar-vector,
and isovector-vector respectively,
which correspond to the exchange of $\sigma$, $\omega$, and $\rho$ mesons in the finite-range
RMF models. We employ the PC-PK1 parametrization proposed in Ref.~\cite{Zhao10},
where the isovector-scalar channel ($\delta$ meson) is neglected since the inclusion
of the isovector-scalar interactions generally does not help improve the description
of ground state properties of stable nuclei.
Attempts to include the isovector-scalar interactions in Ref.~\cite{Burv02}
show that the addition of isovector-scalar terms does not incorporate real physical
improvements and is not required for a viable description of the strong interaction
in finite nuclei.
In order to study the influence of nuclear symmetry energy, we introduce an additional
coupling ($\gamma_{C}$) between the isoscalar-vector and isovector-vector interactions
based on the PC-PK1 parametrization, which plays a crucial role in modifying the density
dependence of nuclear symmetry energy.

In the mean-field approximation, the interactions in the Lagrangian density
are replaced by their expectation values, which can be expressed in terms of corresponding
local densities. The energy density functional of the system is derived from the energy-momentum
tensor. By using standard variational techniques, one can obtain the Dirac equation for the nucleon,
\begin{eqnarray}
\label{DiracEq}
\bigg[-i\mathbf{\alpha}\cdot \mathbf{\nabla}
 &+& \beta\left(m_{N}+V_{S}\right)
    + V_{V} + \tau_{3} V_{TV}  \nonumber \\
 &+& \left. e \frac{1-\tau_3}{2} A
 \right] \psi_k
 =\epsilon_{k}\psi_k,
\end{eqnarray}
where the potentials are given by the following relations:
\begin{eqnarray}
 V_{S}  &=& \alpha_S\rho_S+\beta_S\rho^2_S+\gamma_S\rho^3_S+\delta_S\triangle\rho_S,\\
 V_{V}  &=& \alpha_V\rho_V+\gamma_V \rho_V^3+\delta_V\triangle \rho_V
           +2\gamma_{C}\rho_V \rho_{TV}^2,\\
 V_{TV} &=& \alpha_{TV} \rho_{TV}+\delta_{TV}\triangle \rho_{TV}+2\gamma_{C}\rho_V^2\rho_{TV} ,
\end{eqnarray}
with $\rho_{S}(\bm r)$, $\rho_{V}(\bm r)$, and $\rho_{TV}(\bm r)$ being the local densities
in scalar, vector, and isovector-vector channels, respectively.
$\triangle \rho_S$, $\triangle \rho_V$, and $\triangle \rho_{TV}$ are the corresponding
derivative terms, where $\triangle$ is the Laplace operator.
We use the nuclear physics convention for the isospin, i.e.,
the neutron is associated with $\tau_3 = +1$ and the proton with $\tau_3 = -1$.
In the Thomas-Fermi approximation, the chemical potentials of nucleons are expressed as
\begin{eqnarray}
\label{CPP}
\mu_{p} &=&\sqrt{{k^{p}_{F}}^{2}+{m^{*}_{N}}^{2}}
           +V_{V} - V_{TV} + e A, \\
\label{CPN}
\mu_{n} &=&\sqrt{{k^{n}_{F}}^{2}+{m^{*}_{N}}^{2}}
           +V_{V} + V_{TV},
\end{eqnarray}
where $m^{*}_{N}(\bm r)=m_{N}+V_{S}(\bm r)$ is the effective nucleon mass and $k^{i}_{F}(\bm r)$
is the Fermi momentum.
We emphasize that the chemical potential is spatially constant throughout the whole system,
while other quantities such as various densities depend on the position $\bm r$.
The electrostatic potential $A(\bm r)$ satisfies the Poisson equation
\begin{equation}
\label{EQ:Coul}
 \triangle A(\bm r)=-e\left[\rho^{p}_{V}(\bm r)-\rho^{e}_{V}\right],
\end{equation}
where the electron number density $\rho^{e}_{V}$  is assumed to be uniform in the system for simplicity.

For nonuniform matter at a given average baryon density $\rho_B$ and fixed proton fraction $Y_p$,
the most stable state is the one with the lowest energy. We calculate the energy of a large
cubic cell by performing the three-dimensional integration,
\begin{equation}
  E=\int d^3 r \,\varepsilon(\bm r),
\end{equation}
where the local energy density within the Thomas-Fermi approximation is given by
\begin{eqnarray}
\label{energy}
\varepsilon(\bm r) &=& \sum_{b=p,n}\frac{1}{\pi ^{2}}\int_{0}^{k_{F}^{b}} dk \,k^{2}
\sqrt{k^{2}+m^{\ast 2}_{N}} \nonumber \\
&& -\frac{1}{2} \alpha_{S}\rho_{S}^2
   -\frac{1}{2} \delta_{S}\rho_{S}\Delta\rho_{S}
   -\frac{2}{3} \beta_{S}\rho_{S}^3 \nonumber \\
&& -\frac{3}{4}\gamma_{S}\rho_{S}^4
   +\frac{1}{2} \alpha_{V}\rho_{V}^2
   +\frac{1}{2} \delta_{V}\rho_{V}\Delta\rho_{V}
\nonumber\\
&& +\frac{1}{4} \gamma_{V}\rho_{V}^4
   +\frac{1}{2} \alpha_{TV}\rho_{TV}^2
   +\frac{1}{2} \delta_{TV}\rho_{TV}\Delta\rho_{TV}
\nonumber \\
&&
   +\frac{1}{2}eA\left(\rho^{p}_{V}-\rho^{e}_{V}\right)
   +\gamma_{C} \rho^{2}_{V}\rho^{2}_{TV} \nonumber \\
&& +\frac{1}{\pi ^{2}}\int_{0}^{k_{F}^{e}} dk \,k^{2}
   \sqrt{k^{2}+m^{2}_{e}}.
\end{eqnarray}
We use the iteration method to solve this problem.
In practice, we start with an initial guess for all density distributions $\rho_{S}(\bm r)$,
$\rho_{V}(\bm r)$, $\rho_{TV}(\bm r)$, and the electrostatic potential $A(\bm r)$.
Then, the chemical potentials, $\mu_p$ and $\mu_n$, are respectively determined by given proton
and neutron numbers inside the cell. Once the chemical potentials are achieved, the new Fermi
momentum, $k^{p}_{F}(\bm r)$ and $k^{n}_{F}(\bm r)$, can be obtained from the relations~(\ref{CPP})
and~(\ref{CPN}), which result in new density distributions. Furthermore, new $A(\bm r)$ is
obtained by solving the Poisson equation. This procedure should be iterated
until convergence is achieved.

\section{Parameters}
\label{sec:3}

The parameters of the RMF-PC models are generally determined by fitting to the
ground-state properties of finite nuclei. Several successful RMF-PC parametrizations
have been proposed and widely used in describing various nuclear properties~\cite{Rusn97,Burv02,Niks08,Zhao10,Ring15,Zhao18}.
In the present work, we employ the PC-PK1 parametrization proposed
by Zhao {\it et al.}~\cite{Zhao10}, which was determined by fitting to observables
of 60 selected spherical nuclei, including the binding energies, charge radii,
and empirical pairing gaps. The PC-PK1 parametrization provides satisfactory
description for both spherical and deformed nuclei throughout the nuclear chart~\cite{Zhao18}.
For completeness, we present the PC-PK1 parametrization of the RMF-PC model in Table~\ref{Tabel:PC-PK1}.
With the PC-PK1 parametrization, the predicted saturation properties of nuclear matter
are as follows:
the saturation density $\rho_0=0.153\,\rm{fm}^{-3}$,
energy per nucleon $E_0=-16.12\,\rm{MeV}$,
incompressibility $K=238\,\rm{MeV}$,
symmetry energy $E_{\rm{sym}}=35.6\,\rm{MeV}$,
and the slope of symmetry energy $L=113\,\rm{MeV}$.
Due to the large value of $L$, the PC-PK1 parametrization of the RMF-PC model predicts
rather large radii and tidal deformabilities of neutron stars~\cite{Sun19},
which are disfavored by recent astrophysical observations as discussed in the introduction.
The inclusion of a coupling term between the isoscalar-vector and isovector-vector interactions
can help to reduce the symmetry energy slope, whereas the properties of symmetric nuclear
matter remain unchanged.
\begin{table}[!htbp]
\caption{Coupling constants in the original PC-PK1 parametrization.}
\label{Tabel:PC-PK1}
\begin{ruledtabular}
\begin{tabular}{ccccc}
Coupling constant & Value                    &  Dimension          \\
\hline
 $~\alpha_S$      &$-3.96291\times10^{-4}$   &  ${\rm MeV}^{-2}$   \\

 $~\beta_S$       &$8.6653\times10^{-11}$    &  ${\rm MeV}^{-5}$   \\

 $~\gamma_S$      &$-3.80724\times10^{-17}$  &  ${\rm MeV}^{-8}$   \\

 $~\delta_S$      &$-1.09108\times10^{-10}$  &  ${\rm MeV}^{-4}$   \\

 $~\alpha_V$      &$2.6904\times10^{-4}$     &  ${\rm MeV}^{-2}$   \\

 $~\gamma_V$      &$-3.64219\times10^{-18}$  &  ${\rm MeV}^{-8}$   \\

 $~\delta_V$      &$-4.32619\times10^{-10}$  &  ${\rm MeV}^{-4}$   \\

$~\alpha_{TV}$    &$2.95018\times10^{-5}$    &  ${\rm MeV}^{-2}$   \\

$~\delta_{TV}$    &$-4.11112\times10^{-10}$  &  ${\rm MeV}^{-4}$   \\
$~\gamma_C$       &            $0$           &  ${\rm MeV}^{-8}$   \\
\end{tabular}
\end{ruledtabular}
\end{table}

\begin{table*}[tbp]
\caption{Parameters $\alpha_{TV}$ and $\gamma_{C}$ generated from the PC-PK1 parametrization
for different slope $L$ at saturation density $\rho_0$ with fixed symmetry energy
$E_{\text{sym}}=27.733$ MeV at $\rho_{\rm{fix}}=0.12\, \rm{fm}^{-3}$.
The last line shows the symmetry energy at saturation density. }
\label{tab:L}
\begin{center}
\begin{tabular}{lcccccccccccc}
\hline\hline
$L$ (MeV) & 40    & 50    & 60    & 70    & 80    & 90    & 100  & 113  \\
\hline
$\alpha_{TV}$ ($10^{-5}\,{\rm MeV}^{-2}$) & 4.0168 & 3.8706 & 3.7243 & 3.5780 & 3.4317 & 3.2854 & 3.1392 &2.9502 \\
$\gamma_{C} $ ($10^{-18}\,{\rm MeV}^{-8}$)
&-6.2734&-5.4131&-4.5528&-3.6924&-2.8321&-1.9718&-1.1114& 0 \\
$E_{\rm{sym}}(\rho_0)$ (${\rm MeV})$ & 31.69 & 32.23 & 32.77   & 33.31   & 33.84  & 34.38  & 34.92  & 35.61 \\
\hline\hline
\end{tabular}%
\end{center}
\end{table*}

\begin{figure}[htbp]
\begin{center}
\includegraphics[clip,width=0.43\textwidth]{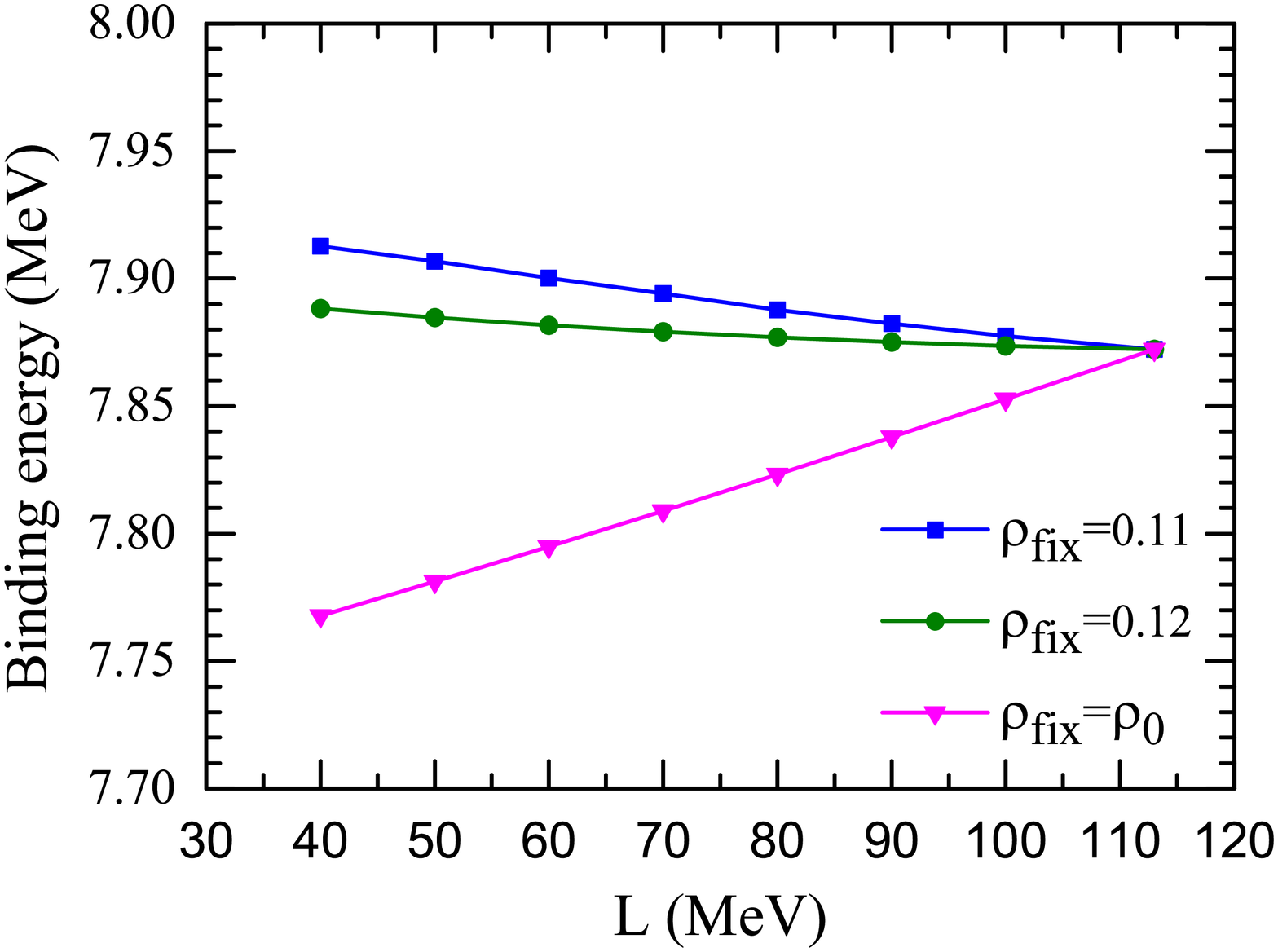}
\caption{Binding energy per nucleon of $^{208}$Pb
vs the symmetry energy slope $L$ with different choices of $\rho_{\rm{fix}}$
based on the PC-PK1 parametrization.}
\label{fig:1Pb208}
\end{center}
\end{figure}

In order to investigate the effect of the symmetry energy slope $L$ on nuclear pasta phases,
we generate a set of RMF-PC models with different values of $L$ at saturation density based on
the PC-PK1 parametrization. For this purpose, we introduce an additional coupling term
with coefficient $\gamma_{C}$ in Eq.~(\ref{LAG}), which corresponds to
the $\omega$-$\rho$ coupling in the finite-range RMF models.
It is well known that this term plays an important role in
modifying the density dependence of the symmetry energy and affecting neutron star
properties~\cite{Mene11,Prov13,IUFSU,Horo01,Horo03,Bao14a,Bao14b,Bao15}.
By simultaneously adjusting the coupling constants $\alpha_{TV}$ and $\gamma_{C}$,
one can achieve a given $L$ at saturation density $\rho_0$ while
keeping the symmetry energy $E_{\rm{sym}}$ fixed at a density of 0.12 fm$^{-3}$.
The resulting parameters $\alpha_{TV}$ and $\gamma_{C}$ are presented in Table~\ref{tab:L}.
The reason for fixing $E_{\rm{sym}}$ at a density of 0.12 fm$^{-3}$ is that
the set of generated models should reproduce similar binding energies of finite nuclei
with the experimental data. It has been found that the binding energy
of finite nuclei is essentially determined by the symmetry energy at a density
of $0.10$--$0.12\, \rm{fm}^{-3}$, not by the one at saturation
density~\cite{Horo01,Mene11,Bao14b,Chen13}.
To examine the sensitivity of the binding energy to the fixed density $\rho_{\rm{fix}}$ of
the symmetry energy, we perform a standard RMF calculation with the PC-PK1 parametrization
for $^{208}$Pb using different choices of $\rho_{\rm{fix}}$.
In Fig.~\ref{fig:1Pb208}, one can see that the binding energy per nucleon of $^{208}$Pb
remains almost unchanged with the variation of $L$ for $\rho_{\rm{fix}}=0.12\, \rm{fm}^{-3}$,
whereas it deviates from the experimental value (7.87 MeV)
when $\rho_{\rm{fix}}=0.11\, \rm{fm}^{-3}$ or $\rho_{\rm{fix}}=\rho_0$ is used.
We note that the result for $L=113$ MeV corresponds to the original PC-PK1 case
with $\gamma_{C}=0$, and therefore all three lines terminate at the same point.

\begin{figure}[htbp]
\begin{center}
\includegraphics[clip,width=0.43\textwidth]{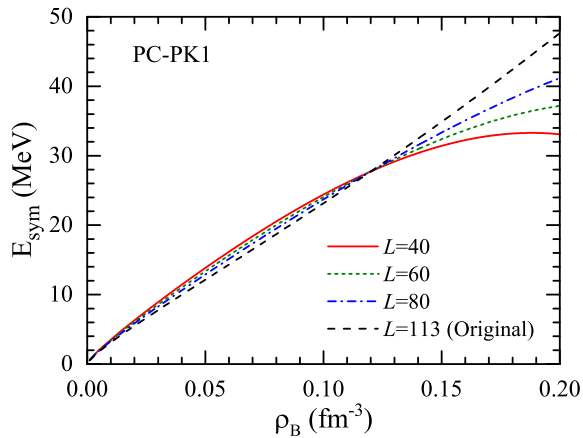}
\caption{Symmetry energy $E_{\rm sym}$ as a function of the
baryon density $\rho_{\rm{B}}$ for the generated models with different $L$.
The symmetry energy is fixed at a density of 0.12 fm$^{-3}$.}
\label{fig:2Esym}
\end{center}
\end{figure}

\begin{figure*}[htbp]
\begin{center}
\includegraphics[clip,width=0.86\textwidth]{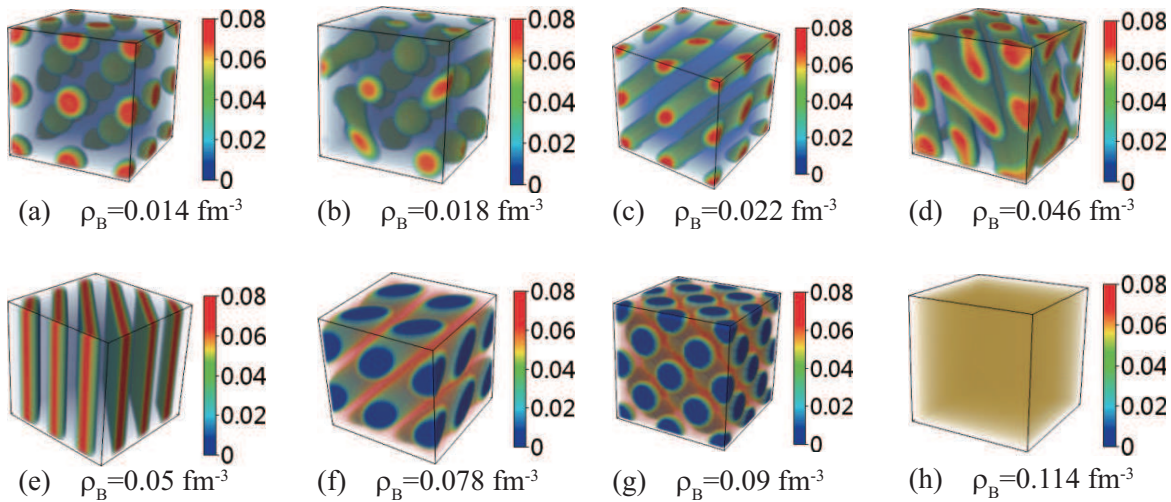}
\caption{Proton density distributions in pasta phases for $Y_p=0.5$
obtained using the RMF-PC model with $L=40$ MeV.}
\label{fig:3YP05}
\end{center}
\end{figure*}

\begin{figure}[htbp]
\begin{center}
\includegraphics[clip,width=0.43\textwidth]{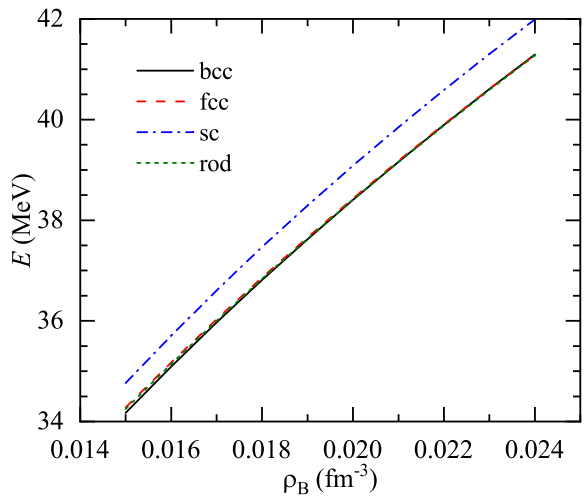}
\caption{Energy per nucleon $E$ for different configurations
observed around the transition from droplets to rods.}
\label{fig:4BCC}
\end{center}
\end{figure}

\begin{figure}[htbp]
\begin{center}
\includegraphics[clip,width=0.43\textwidth]{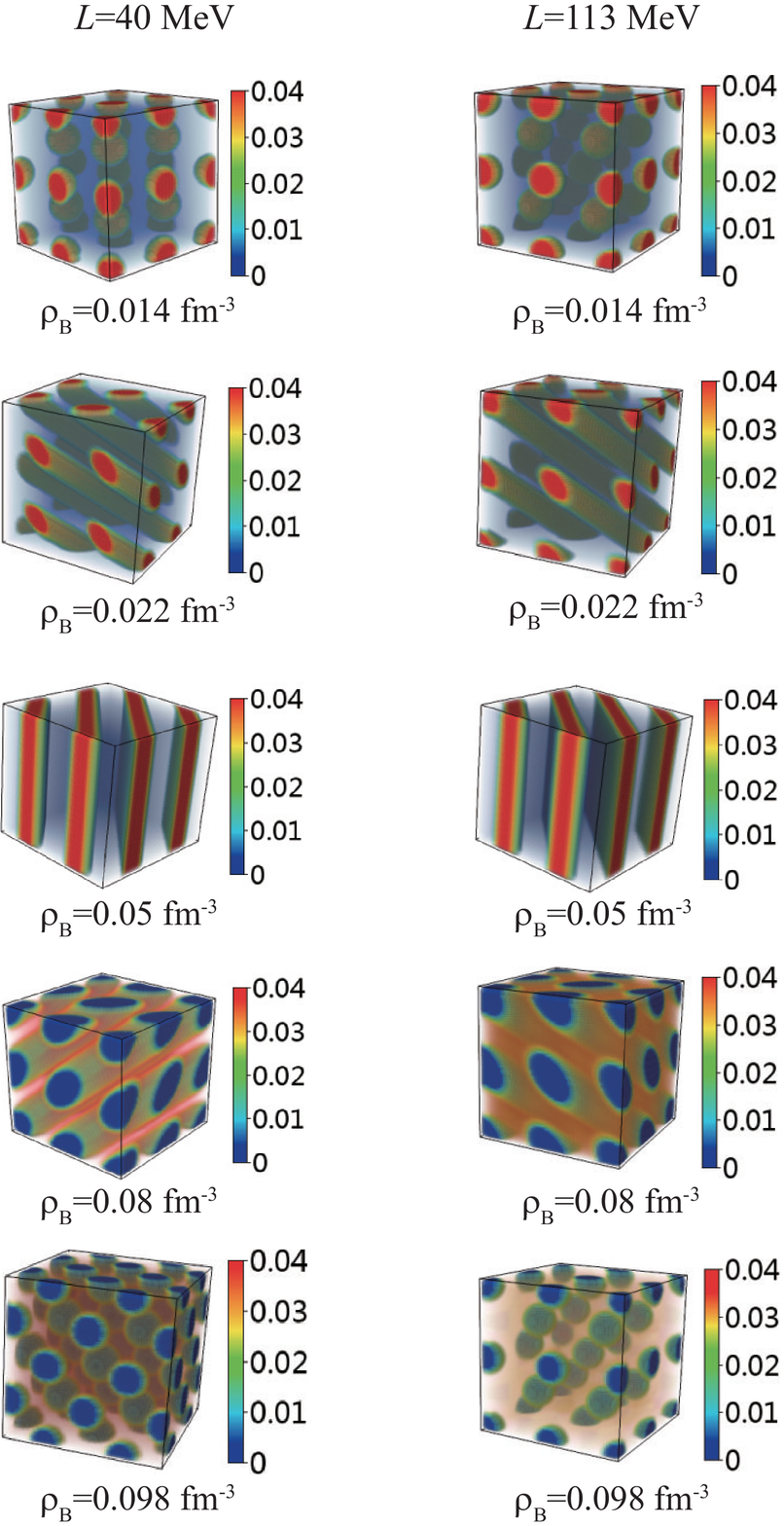}
\caption{Proton density distributions in typical pasta phases
at a fixed proton fraction of $Y_p=0.3$.
The results obtained with $L=40$ MeV (left panels) are compared
to those with $L=113$ MeV (right panels).}
\label{fig:5YP03}
\end{center}
\end{figure}

\begin{figure}[htbp]
\begin{center}
\includegraphics[clip,width=0.43\textwidth]{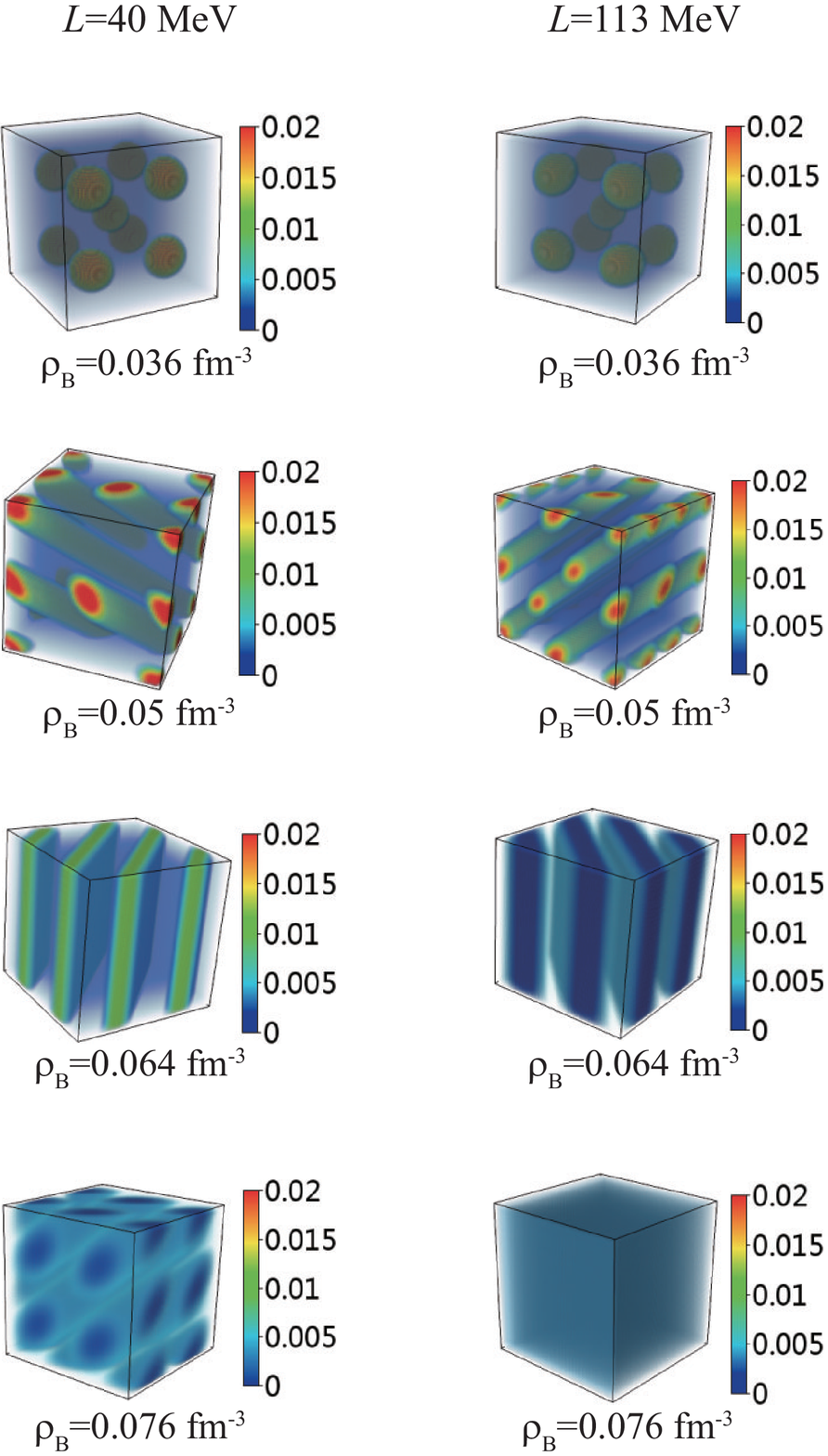}
\caption{Proton density distributions in typical pasta phases
at a fixed proton fraction of $Y_p=0.05$.
The results obtained with $L=40$ MeV (left panels) are compared
to those with $L=113$ MeV (right panels).}
\label{fig:6YP005}
\end{center}
\end{figure}

We emphasize that all generated models in Table~\ref{tab:L} have the same isoscalar
properties and fixed symmetry energy at $\rho_B=0.12\, \rm{fm}^{-3}$,
but they have different density dependence of the symmetry energy.
In Fig.~\ref{fig:2Esym}, we display the symmetry energy $E_{\rm{sym}}$ as a function
of the baryon density $\rho_B$ for the set of generated models.
It is obvious that all models have the same $E_{\rm{sym}}$ at a density of 0.12 fm$^{-3}$,
but they have different $E_{\rm{sym}}$ at lower and higher densities
due to the difference of the slope $L$.
One can see that a smaller $L$ corresponds to a larger (smaller)
$E_{\rm{sym}}$ at lower (higher) densities.
It is feasible and interesting to use these models for studying
the effect of $L$ on pasta structures at subnuclear densities.

In the present work, we perform three-dimensional calculations in a large cubic cell
with the periodic boundary condition. Considering the balance between desired accuracy
and computational time, we use the cell size of $60\,{\rm fm}$ with $64$ grid points
in each direction. Such choices are generally large enough for obtaining convergent
results in three-dimensional calculations of nuclear pasta~\cite{Okam12,Sage16}.
It was reported in Ref.~\cite{Sage16} that changing the box size from $24\,{\rm fm}$
to $48\,{\rm fm}$ in the Skyrme Hartree-Fock calculation would not significantly
change the total energies of the ground state.
In our calculations, enlarging the number of grid points from 64 to 128 leads to a
energy difference within a few keV, which is negligible for determining the pasta configuration.
The computational time in three-dimensional calculations is estimated to scale
as $n^3$, where $n$ is the number of grid points in each direction.
Therefore, the calculation with a larger $n$ is much more time-consuming.
Using the cell size of $60\,{\rm fm}$ with $n=64$, the grid spacing is $0.9375\,{\rm fm}$,
which is a reasonable value for three-dimensional calculations of nuclear pasta~\cite{Okam12,Sage16}.
At a typical density of $\rho_B = 0.05\, \rm{fm}^{-3}$, there are about 10800 nucleons
in a cubic cell with the size of $60\,{\rm fm}$, where several periods of pasta structures can be
formed (see Figs. 3, 5, and 6 below).
\section{Results and discussion}
\label{sec:4}

In this section, we present the results of three-dimensional calculations for nonuniform
matter at subnuclear densities. We explore the pasta structures for both cold stellar matter with a fixed proton fraction and neutron-star matter in $\beta$ equilibrium.
The influence of nuclear symmetry energy is examined by using the generated models
from the PC-PK1 parametrization.

\subsection{Pasta structures with a fixed proton fraction}
\label{sec:4.1}

We first present and compare the pasta structures for fixed proton fraction
by using two limit RMF-PC models in Table~\ref{tab:L}, namely the ones with $L=40$ and 113 MeV.
In symmetric nuclear matter ($Y_p=0.5$), these two models provide almost the same
features of pasta structures.
In Fig.~\ref{fig:3YP05}, we show the proton density distributions of nonuniform
matter for $Y_p=0.5$ in a cubic cell with a length of $60\,{\rm fm}$.
The results are obtained from the three-dimensional Thomas-Fermi calculations by
using the RMF-PC model with $L=40$ MeV. At a low density of $\rho_B=0.014\,\rm{fm}^{-3}$,
the matter forms a crystalline structure of droplets.
As the density $\rho_B$ increases, typical pasta phases like rods, slabs, tubes, and bubbles
are observed before the transition to uniform matter.
In addition, some intermediate structures around the shape transition are also
observed [see Fig.~\ref{fig:3YP05}(b) and~\ref{fig:3YP05}(d)], which make the transition
between different shapes more smooth.
In our calculations, the ground state at low densities is a body-centered cubic (bcc)
lattice of droplets, whereas a face-centered cubic (fcc) lattice may appear as a
metastable state. This is consistent with the previous studies in Refs.~\cite{Oyam93,Wata03},
but inconsistent with the results in Refs.~\cite{Okam12,Okam13} where the fcc lattice
is energetically more favorable than the bcc one.
In practical calculations, the final configurations are somewhat influenced by
the initial density distributions. When different initial configurations are used,
some metastable states may arrive after the convergence is achieved.
In Fig.~\ref{fig:4BCC}, we compare the energy per nucleon $E$ among different
configurations observed around the transition from droplets to rods.
It is found that a simple cubic (sc) lattice of droplets emerges as a metastable state
at low densities, whose energy is obviously larger than that of the bcc lattice.
On the other hand, the energy of an fcc lattice is only slightly higher than that in
the bcc case, while their energy difference decreases as the density increases.
At the density $\rho_B > 0.02\,\rm{fm}^{-3}$, the rod phase becomes the ground state
with the lowest energy, but its energy per nucleon $E$ is only a few keV lower than
that of the bcc lattice. In Fig.~\ref{fig:3YP05}(g), an fcc lattice of bubbles is observed
before the transition to uniform matter, which is consistent with the results in Refs.~\cite{Okam12,Okam13}.

\begin{figure}[htbp]
\begin{center}
\includegraphics[clip,width=0.43\textwidth]{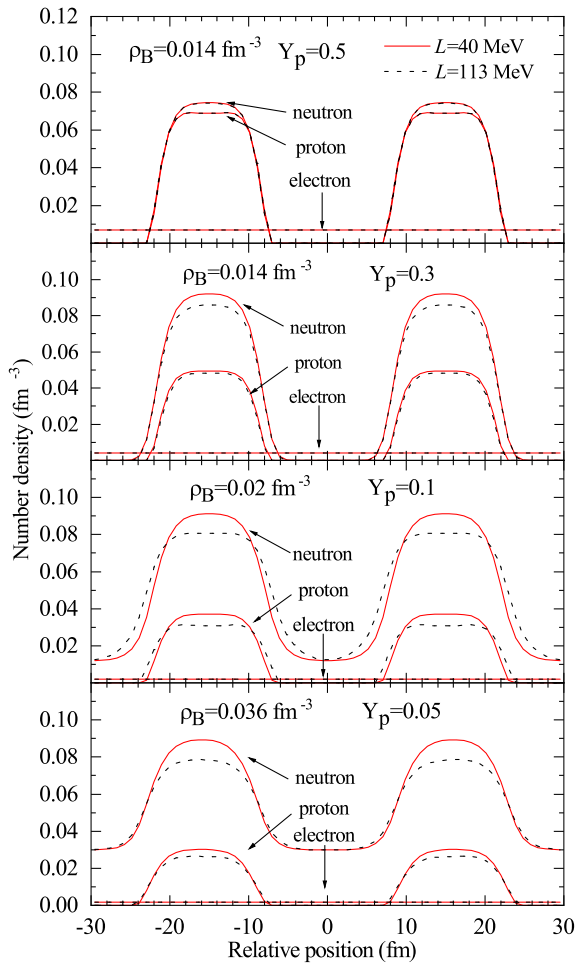}
\caption{Density distributions of protons, neutrons, and electrons
along a line passing through the center of the droplets.
The results obtained with $L=40$ MeV (solid lines) are compared to those
with $L=113$ MeV (dashed lines).}
\label{fig:7rhonpe}
\end{center}
\end{figure}

To explore the influence of symmetry energy and its slope, we compare the pasta
structures obtained using the models with $L=40$ and 113 MeV for lower values
of $Y_p$, where the isovector part is expected to play a crucial role.
In Figs.~\ref{fig:5YP03} and~\ref{fig:6YP005}, we display the proton density distributions
in typical pasta phases for $Y_p=0.3$ and 0.05, respectively.
It is found that the pasta structures obtained with $L=40$ MeV (left panels) and
$L=113$ MeV (right panels) show similar features in the case of $Y_p=0.3$,
but significant differences are observed for a low value of $Y_p=0.05$.
In Fig.~\ref{fig:6YP005}, one can see that the proton densities at the center of nuclear
pastas obtained with $L=40$ MeV is relatively larger than that of $L=113$ MeV.
Furthermore, at a density of $\rho_B=0.076\,\rm{fm}^{-3}$,
the matter forms a crystalline structure of bubbles with $L=40$ MeV,
but it is already in uniform phase with $L=113$ MeV.
The differences of pasta properties between $L=40$ and 113 MeV can be seen more clearly
in Fig.~\ref{fig:7rhonpe}, where the density profiles in droplet configurations for
different $Y_p$ are displayed along a line passing through the center of the droplets.
In the top panel with $Y_p=0.5$, there is no visible difference in the density distributions
between the two models. With decreasing $Y_p$, one can see that the model with $L=40$ MeV
results in larger neutron densities at the center of the droplet compared to that
with $L=113$ MeV, and this trend is more pronounced for lower values of $Y_p$.
Meanwhile, the proton densities at the center of the droplets with $L=40$ MeV are
only slightly higher than those with $L=113$ MeV.
This behavior can be understood from the density dependence of
the symmetry energy $E_{\rm{sym}}$ shown in Fig.~\ref{fig:2Esym}.
With a smaller slope $L=40$ MeV, $E_{\rm{sym}}$ is relatively small at higher
densities ($\rho_B>0.12\, \rm{fm}^{-3}$), which leads to larger neutron densities
at the center of the droplet. Similar differences between $L=40$ and 113 MeV
are also observed in other pasta configurations.
On the other hand, we can see that dripped neutrons exist outside the droplets for
small values of $Y_p=0.1$ and 0.05, whereas all nucleons participate in forming
nuclear clusters for $Y_p=0.3$ and 0.5. Generally, a free neutron gas may appear
for $Y_p<0.3$ and its density increases with decreasing $Y_p$.

\begin{figure}[htbp]
\begin{center}
\includegraphics[clip,width=0.43\textwidth]{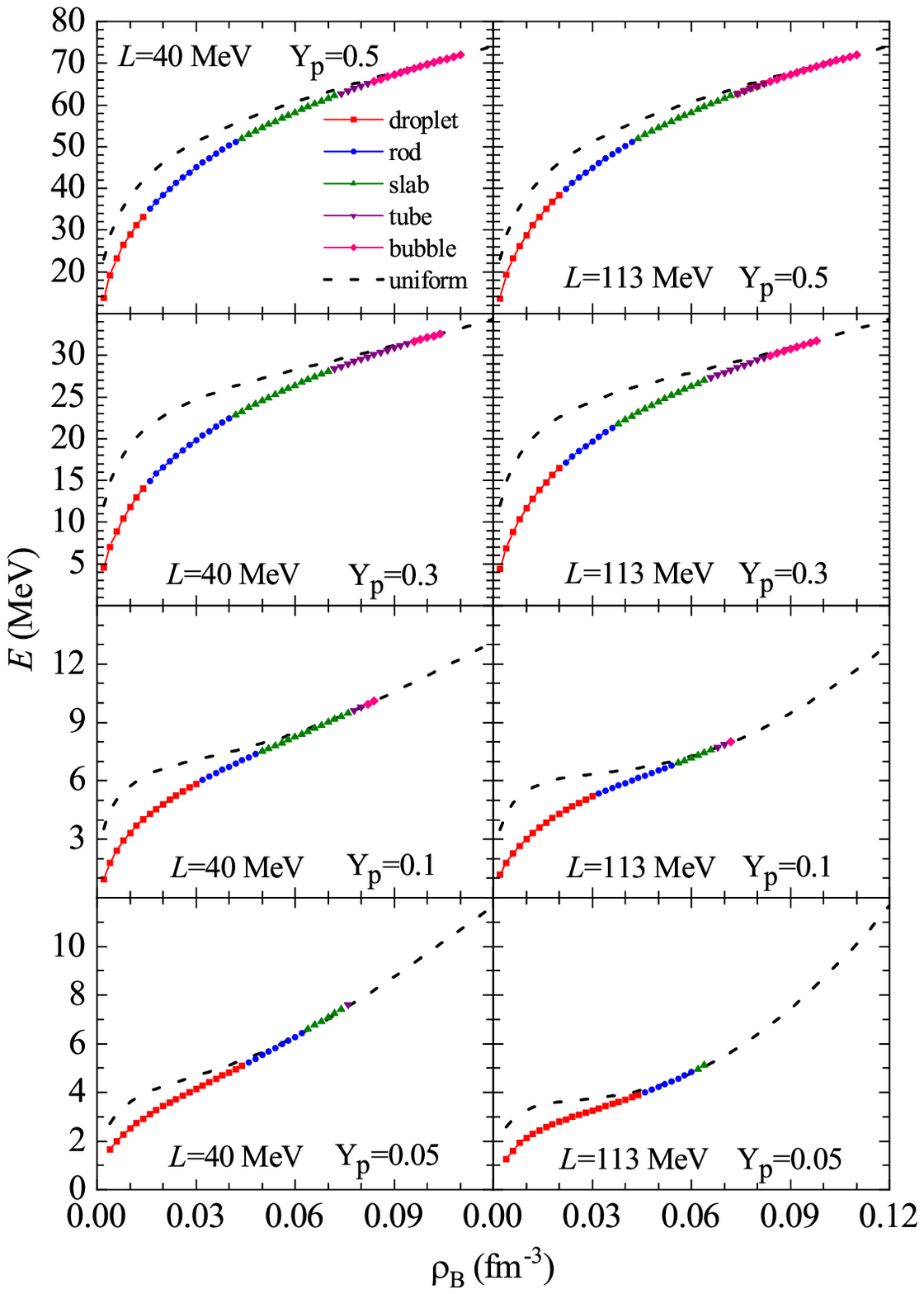}
\caption{Energy per nucleon $E$ as a function of the baryon
density $\rho_B$ for $Y_p=0.5$, 0.3, 0.1, and 0.05 using the models
with $L=40$ MeV (left panels) and $L=113$ MeV (right panels).
For comparison, the results of uniform matter are also displayed by dashed lines.}
\label{fig:8E}
\end{center}
\end{figure}

In Fig.~\ref{fig:8E}, we show the energy per nucleon $E$ as a function of the baryon
density $\rho_B$ for $Y_p=0.5$, 0.3, 0.1, and 0.05.
For comparison, the results of uniform matter are displayed by dashed lines,
which are obviously higher than those of pasta phases at lower densities.
The results obtained with $L=40$ MeV (left panels) are compared to those
with $L=113$ MeV (right panels).
One can see that the behaviors of $E$ are very similar between these two models
for larger values of $Y_p=0.5$ and 0.3, whereas significant differences are
observed for $Y_p=0.1$ and 0.05.
The model with $L=40$ MeV predicts relatively large $E$ and late transition
to uniform matter compared to that with $L=113$ MeV.
This is because $E_{\rm{sym}}$ in the model with $L=40$ MeV is larger than
that with $L=113$ MeV at low densities (see Fig.~\ref{fig:2Esym}).
It is seen that the transition from pasta phase to uniform matter occurs
at lower densities for smaller values of $Y_p$, and some pasta shapes like the
bubble configuration could not appear before the transition to uniform matter
in the case of $Y_p=0.05$.

\subsection{Inner crust of neutron stars}
\label{sec:4.2}

\begin{figure}[htbp]
 \begin{center}
  \includegraphics[clip,width=0.43\textwidth]{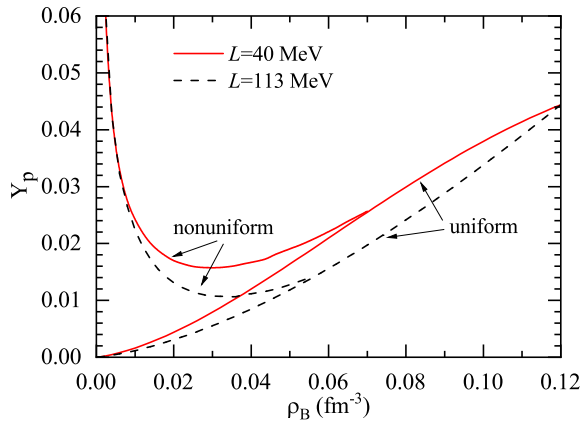}
  \caption{Proton fractions as a function of the baryon density in
  neutron-star matter with nonuniform and uniform distributions.}
  \label{fig:9Yprho}
 \end{center}
\end{figure}

\begin{figure}[htbp]
 \begin{center}
  \includegraphics[clip,width=0.43\textwidth]{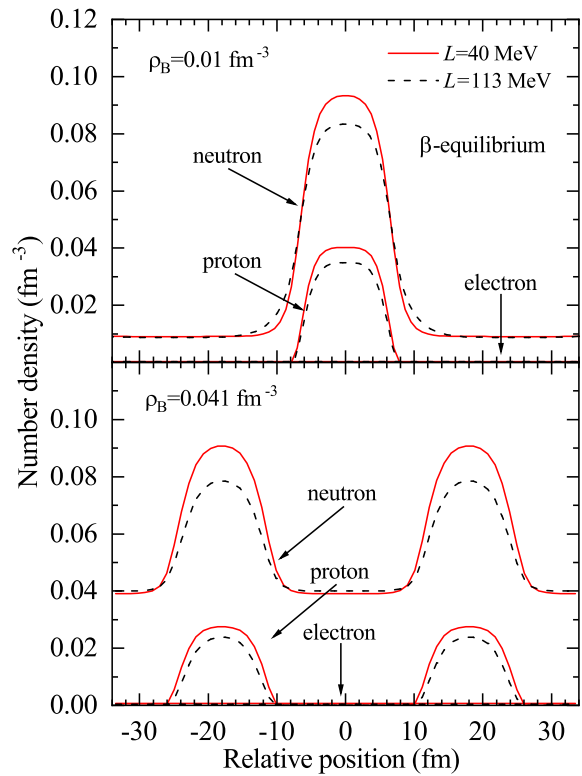}
  \caption{Density distributions of protons, neutrons, and electrons
  in $\beta$ equilibrium along a line passing through the center of the droplets.
  The results obtained with $L=40$ MeV are compared to those of $L=113$ MeV.}
  \label{fig:10rhobeta}
 \end{center}
\end{figure}

\begin{figure*}[htbp]
 \begin{center}
\begin{tabular}{cc}
  \includegraphics[clip,width=8.6cm]{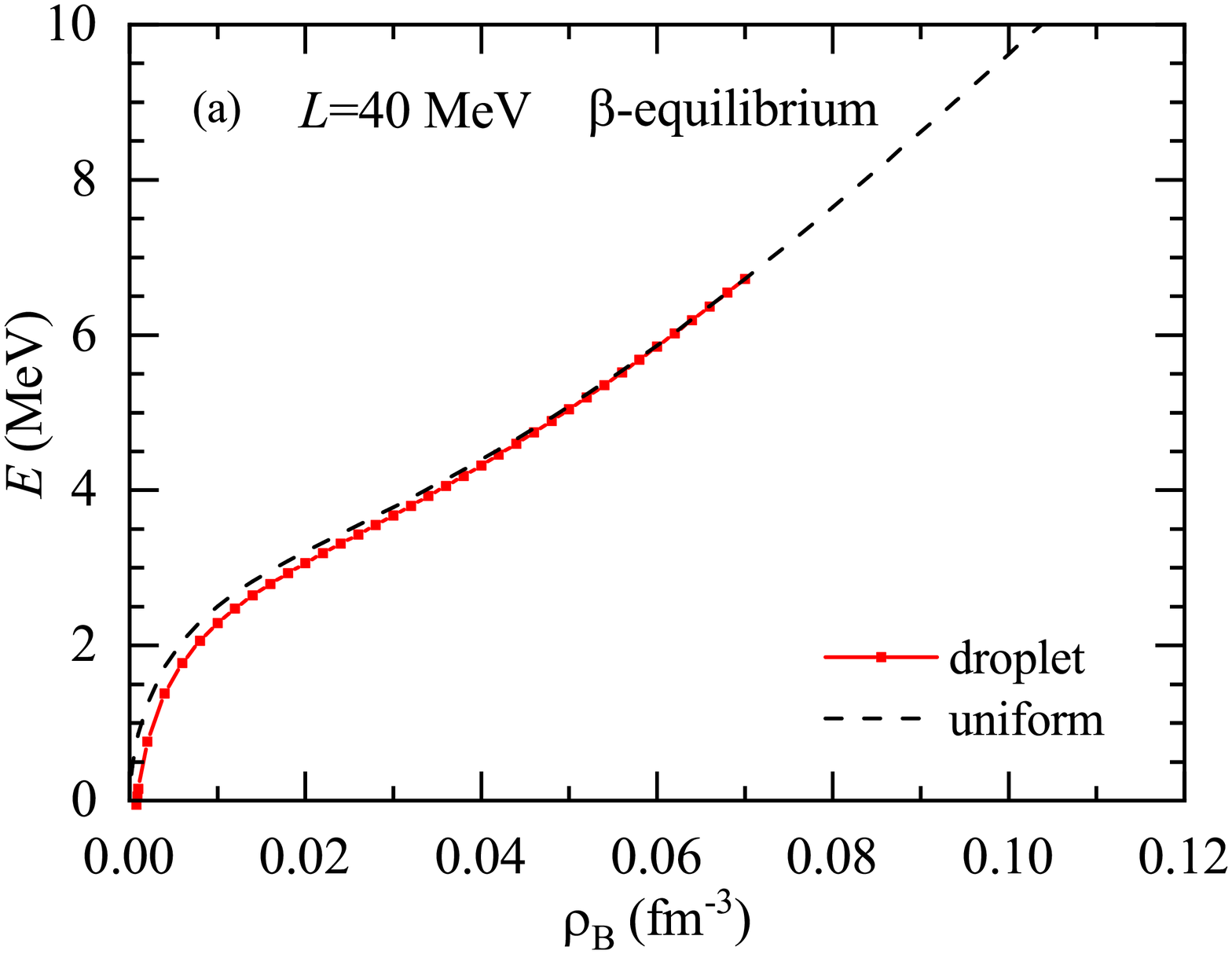} &
  \includegraphics[clip,width=8.6cm]{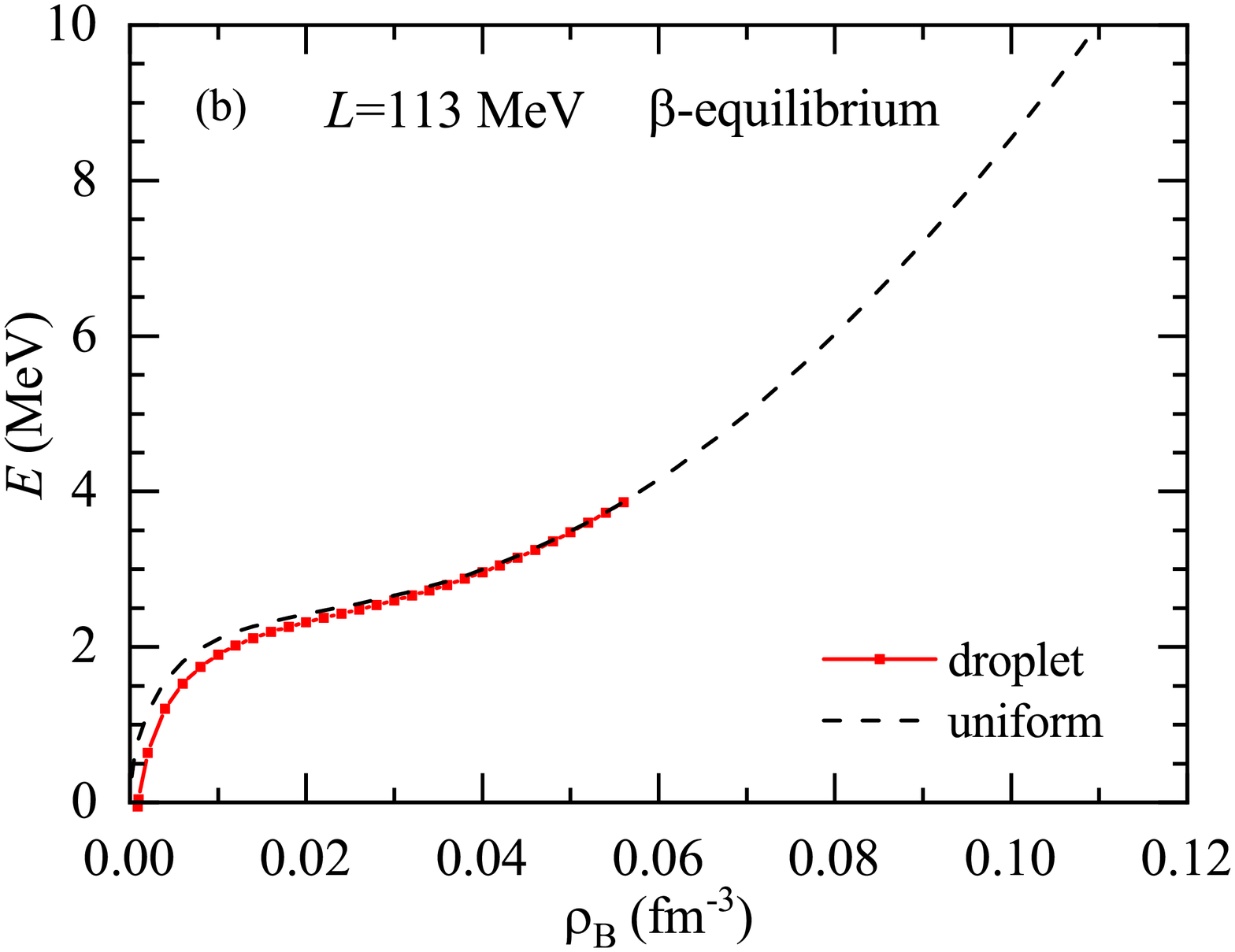} \\
\end{tabular}
  \caption{Energy per nucleon $E$ as a function of the baryon
  density $\rho_B$ using the models with $L=40$ MeV (a) and $L=113$ MeV (b).
  For comparison, the results of uniform matter are also displayed by dashed lines.}
  \label{fig:11Ebeta}
 \end{center}
\end{figure*}

\begin{figure*}[htbp]
 \begin{center}
\begin{tabular}{cc}
  \includegraphics[clip,width=8.6cm]{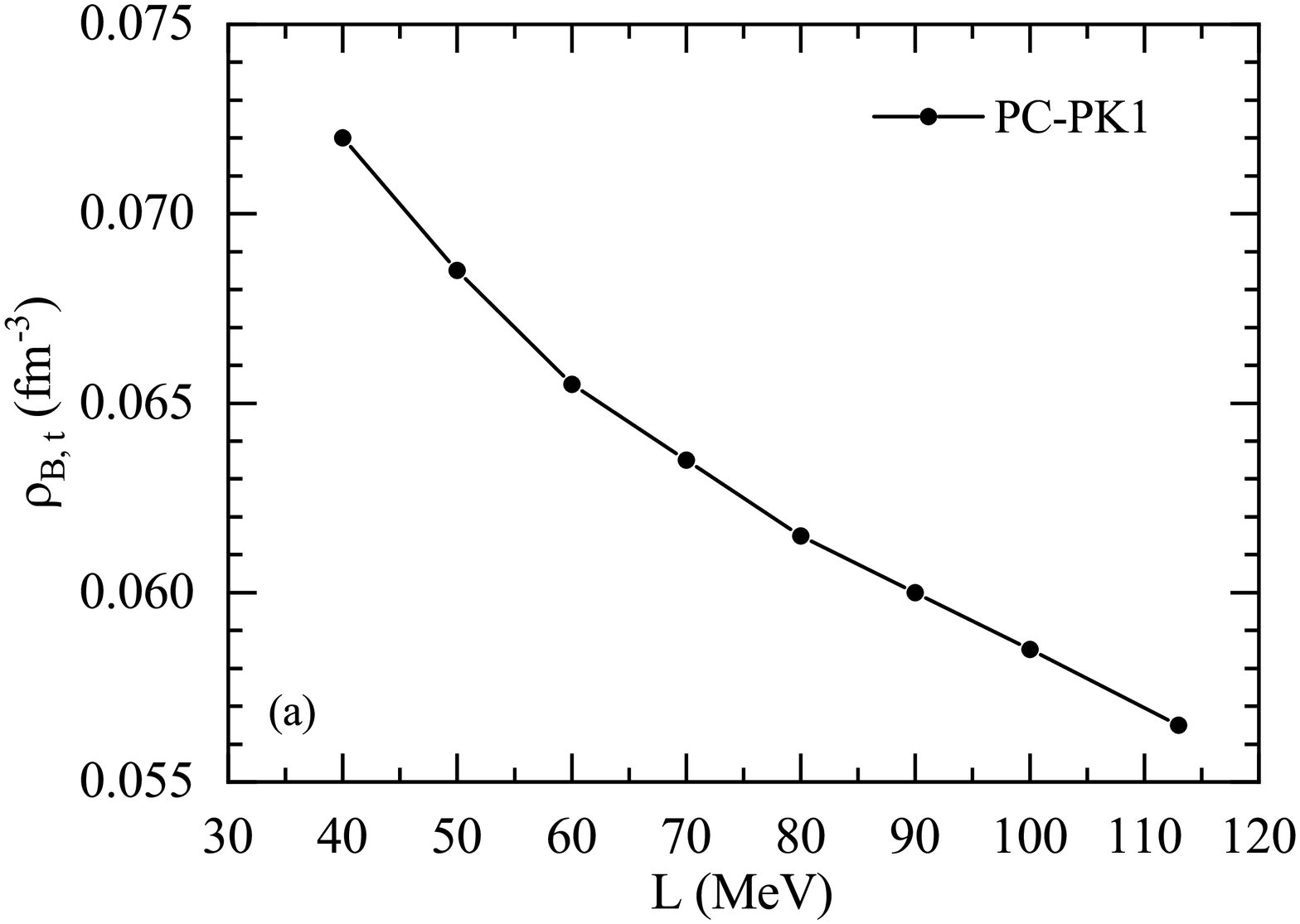} &
  \includegraphics[clip,width=8.6cm]{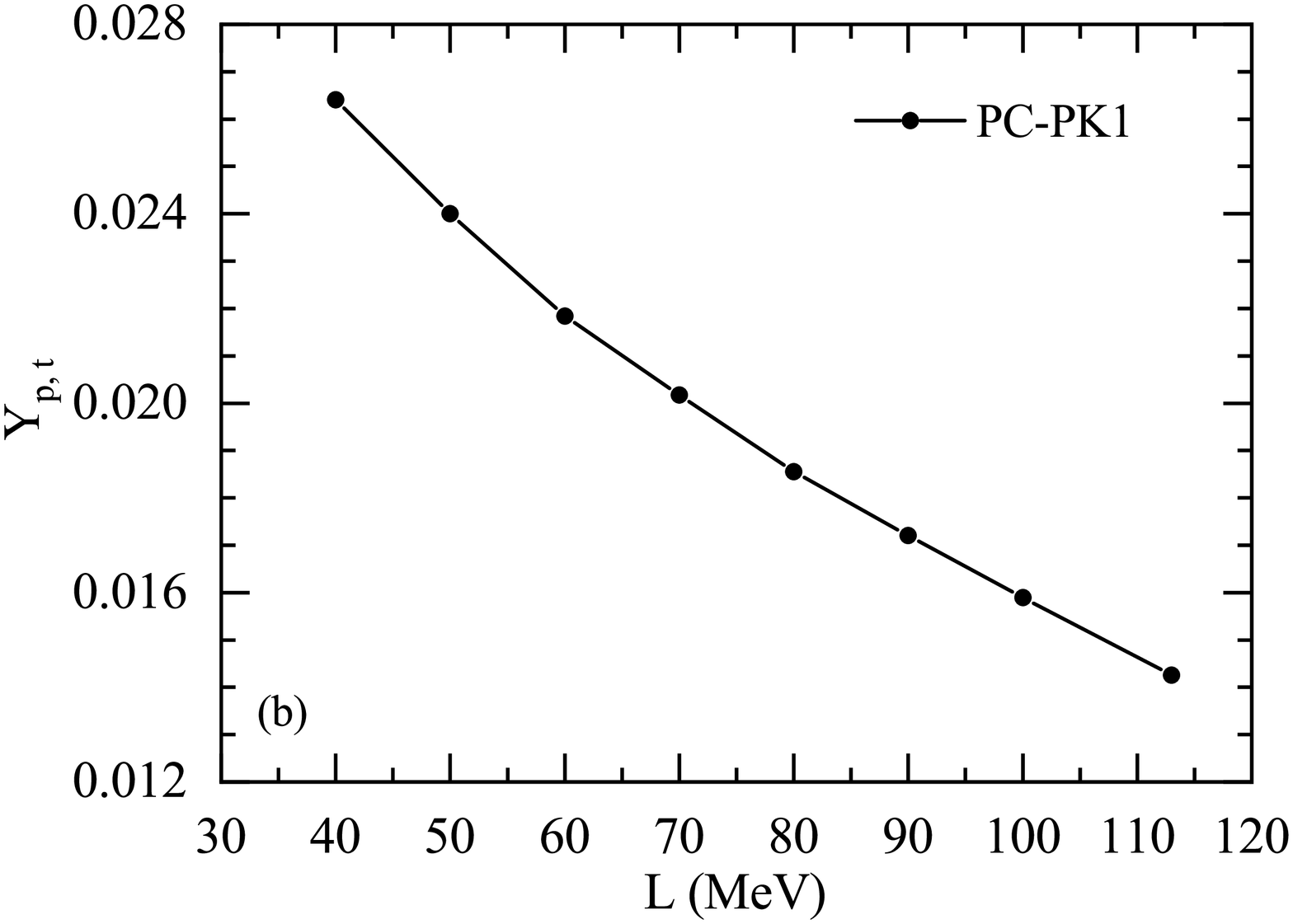} \\
\end{tabular}
  \caption{Crust-core transition density $\rho_{B,t}$ (a) and proton fraction
  at the transition point $Y_{p,t}$ (b) as a function of the symmetry energy slope $L$
  using the generated models based on the PC-PK1 parametrization. }
  \label{fig:12TL}
 \end{center}
\end{figure*}


To describe nonuniform matter in the inner crust of neutron stars,
we perform fully three-dimensional calculations in a cubic box with
periodic boundary conditions, where the conditions of $\beta$ equilibrium
and charge neutrality are satisfied.
We carry out the calculations using the models with $L=40$ and 113 MeV,
so as to examine the influence of the symmetry energy slop $L$.
The matter in neutron-star crusts contains protons, neutrons, and electrons,
where the proton fraction $Y_p$ is determined by the $\beta$ equilibrium
condition and its value is sensitive to the behavior of the symmetry energy.
In Fig.~\ref{fig:9Yprho}, we display the proton fraction $Y_p$ of nonuniform
matter in neutron-star crusts as a function of the baryon density $\rho_B$
using the models with $L=40$ and 113 MeV, where the results of uniform matter
are also shown for comparison.
It is found that both models predict rather small values of $Y_p$ in the
density region of $0.02 < \rho_B < 0.12\, \rm{fm}^{-3}$, where nonspherical
pasta structures are expected to appear. However, only spherical droplets are
observed in our calculations before the crust-core transition which occurs
at $\rho_B \simeq 0.072\, \rm{fm}^{-3}$ with $L=40$ MeV and
at $\rho_B \simeq 0.057\, \rm{fm}^{-3}$ with $L=113$ MeV.
Due to smaller values of $Y_p$ obtained in $\beta$ equilibrium, it is unlikely
to form nonspherical pasta in neutron-star crusts, and meanwhile the transition
to uniform matter occurs at lower densities. This is in contrast to the results
with a fixed $Y_p$ shown in the previous subsection.
One can see that at low densities, $Y_p$ of nonuniform matter is significantly
larger than that of uniform matter. This is because the formation of nuclear
clusters can largely reduce the chemical potential of protons, which leads to
an enhancement of $Y_p$ in nonuniform matter.
Comparing the results between $L=40$ and $113$ MeV,
we see that a smaller $L$ corresponds to a larger $Y_p$ in
both nonuniform and uniform cases.
This correlation can be understood from the density dependence of
the symmetry energy $E_{\rm{sym}}$ shown in Fig.~\ref{fig:2Esym}.
At low densities ($\rho_B<0.12\, \rm{fm}^{-3}$), the model with $L=40$ MeV
has larger $E_{\rm{sym}}$ than that with $L=113$ MeV, and as a result,
it favors to contain more protons in the system.

In Fig.~\ref{fig:10rhobeta}, we plot the density distributions of protons, neutrons,
and electrons in $\beta$ equilibrium along a line passing through the center of the
droplets. At a low density of $\rho_B=0.01\, \rm{fm}^{-3}$ (top panel),
nuclear clusters are formed in a dripped neutron gas with large space between the clusters.
At $\rho_B=0.041\, \rm{fm}^{-3}$ (bottom panel), the distance between droplets becomes
relatively small and the dripped neutrons are significantly enhanced.
One can see that there are clear differences between the results of $L=40$
and $113$ MeV. The model with $L=40$ MeV yields larger neutron and proton densities
at the center of droplets compared to those with $L=113$ MeV, and the difference is more
pronounced for neutrons. This effect is caused by different density dependence of
the symmetry energy $E_{\rm{sym}}$ (see Fig.~\ref{fig:2Esym}),
where a smaller $E_{\rm{sym}}$ at $\rho_B>0.12\, \rm{fm}^{-3}$ with $L=40$ MeV
leads to larger neutron densities inside nuclear clusters.
In Fig.~\ref{fig:11Ebeta}, we display the energy per nucleon $E$ as a function of
the baryon density $\rho_B$ for neutron-star matter in $\beta$ equilibrium.
Compared to the case with fixed $Y_p$ in Fig.~\ref{fig:8E}, the reduction of $E$
in nonuniform matter is less pronounced. This is because $Y_p$ in $\beta$ equilibrium
is rather small (see Fig.~\ref{fig:9Yprho}), that only small fraction of nucleons
can form clusters which do not affect the total energy very much.
It is shown that $E$ obtained with $L=40$ MeV is higher than that with
$L=113$ MeV, since the model with $L=40$ MeV has larger symmetry energy
$E_{\rm{sym}}$ at low densities.

To study the correlation between the symmetry energy slope $L$ and the
crust-core transition, we perform calculations for nonuniform matter in
$\beta$ equilibrium by employing the set of generated models given in Table~\ref{tab:L}.
We display in Fig.~\ref{fig:12TL} the crust-core transition density $\rho_{B,t}$
and proton fraction at the transition point $Y_{p,t}$ as a function
of $L$ using the generated models based on the PC-PK1 parametrization.
It is shown that both $\rho_{B,t}$ and $Y_{p,t}$ decrease with increasing $L$.
These correlations are consistent with those reported in Refs.~\cite{Bao15,Oyam07,Duco10,Duco08,Pais16a,Pais16b}.
In the present work, we obtain $\rho_{B,t}=0.072\, \rm{fm}^{-3}$ for $L=40$ MeV,
and it decreases to $0.057\, \rm{fm}^{-3}$ for $L=113$ MeV.
The correlation between $\rho_{B,t}$ and $L$ can be understood from
an analysis in the liquid-drop model~\cite{Duco10}, where the energy-density
curvature of pure neutron matter at saturation density is approximately proportional
to $L$. The crust-core transition occurs when the energy-density curvature
becomes negative, i.e., spinodal instability. This implies that a larger $L$
requires a lower $\rho_{B,t}$ for reaching the negative curvature region.
On the other hand, the decrease of $Y_{p,t}$ is related to the density
dependence of $E_{\rm{sym}}$. The model with a larger $L$ has a smaller
$E_{\rm{sym}}$ at $\rho_B<0.12\, \rm{fm}^{-3}$, so it results in a smaller
$Y_{p,t}$ at the transition point. It is noteworthy that the crust-core
transition depends on both nuclear interaction and description of
nonuniform matter.

\subsection{Properties of neutron stars }
\label{sec:4.3}

\begin{figure}[htbp]
 \begin{center}
  \includegraphics[clip,width=0.43\textwidth]{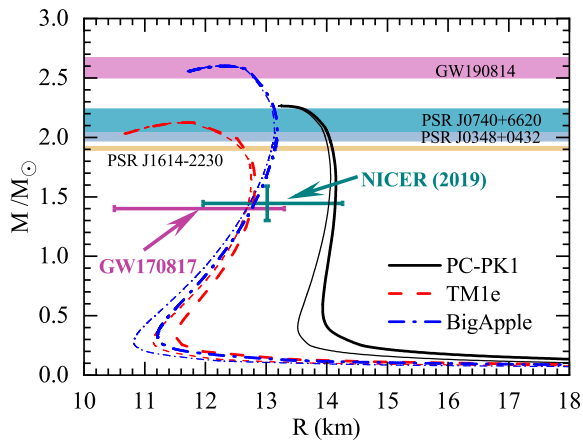}
  \caption{Mass-radius relations of neutron stars by using different combinations
  of the core and crust segments. The results using the inner crust EOS with $L$=40
  and 113 MeV are shown by thick and thin lines, respectively.
  The colored horizontal bars indicate the mass measurements of
  PSR J1614--2230~\cite{Demo10,Fons16,Arzo18}, PSR J0348+0432~\cite{Anto13},
  and PSR J0740+6620~\cite{Crom19}, while the mass constraint from
  GW190814~\cite{Abbo190814} is shown in pink color.
  The constraints from NICER~\cite{Mill19} and GW170817~\cite{Abbo18} are also indicated. }
  \label{fig:13MR}
 \end{center}
\end{figure}

\begin{figure}[htbp]
 \begin{center}
  \includegraphics[clip,width=0.43\textwidth]{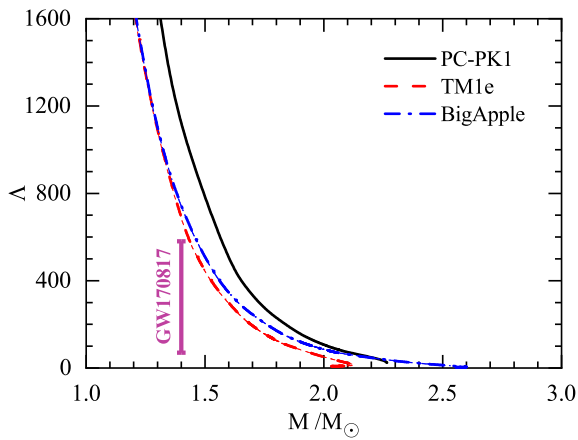}
  \caption{Dimensionless tidal deformability $\Lambda$ as a function of the neutron-star mass $M$.
  The vertical line represents the constraints on $\Lambda_{1.4}$ from the analysis of GW170817~\cite{Abbo18}.}
  \label{fig:14LAMB}
 \end{center}
\end{figure}

The properties of neutron stars, such as mass-radius relations and tidal
deformabilities, can be obtained by solving the Tolman-Oppenheimer-Volkoff (TOV)
equation using the EOS over a wide range of densities.
Generally, the EOS is composed of three segments: the outer crust, the inner crust,
and the liquid core. In the present work, we focus on nonuniform structure of the
inner crust. It is interesting to explore the possible influence from the inner crust
on various properties of neutron stars.
In practical calculations, we adopt the Baym--Pethick--Sutherland (BPS) EOS~\cite{BPS71}
for the outer crust below the neutron drip density, while the EOS obtained from
our three-dimensional calculations is adopted for the inner crust.
It is known that the mass-radius relations of neutron stars are dominantly determined
by the core EOS at high densities, where large uncertainties exist
among different models. Considering observational constraints on neutron-star masses
and radii, we employ several core EOSs, namely PC-PK1~\cite{Sun19}, TM1e~\cite{Shen20},
and BigApple~\cite{Fatt20}, which are matched to the inner crust segments
around the transition density. The PC-PK1 parametrization of the RMF-PC
model used in our calculations of nonuniform matter has a large symmetry
energy slope $L=113$ MeV, which predicts rather large radii
and tidal deformabilities of neutron stars as reported in Ref.~\cite{Sun19}.
Furthermore, the contribution from $\gamma_{V}$ relevant term in Eq.~(\ref{LAG})
may result in negative pressures at high densities, which leads to the difficulty
of reaching the maximum mass of neutron stars.
The TM1e and BigApple parametrizations in the RMF approach have relatively small
slope parameter $L\simeq 40$ MeV, which are more consistent with current
constraints from astrophysical observations.
The TM1e parametrization has been successfully used to construct the EOS for
numerical simulations of core-collapse supernovae~\cite{Shen20}.
The BigApple parametrization has been proposed to account for a $2.6 M_\odot$
compact object observed in GW190814~\cite{Fatt20}.
We note that all of the models can provide satisfactory descriptions of
finite nuclei, and meanwhile they can satisfy the $2 M_\odot$ constraint for neutron stars.

To examine the influence of the inner crust, we adopt the EOS of nonuniform matter
based on the RMF-PC models with $L=40$ and 113 MeV, as described in the previous section.
In Fig.~\ref{fig:13MR}, we display the resulting mass-radius relations of neutron stars
by using different combinations of the core and crust segments.
The results using the inner crust EOS with $L=40$ and 113 MeV are shown by thick and
thin lines, respectively. It is found that massive neutron stars are insensitive
to the inner crust EOS, while visible differences are observed in low-mass neutron stars.
For the same core EOS, the radii of neutron stars obtained using the inner crust EOS
with $L=40$ MeV are slightly larger than that with $L=113$ MeV.
On the other hand, the differences caused by the core EOS are much more pronounced.
In the case with the PC-PK1 core EOS, its large symmetry energy slope $L=113$ MeV
results in the radius of a canonical $1.4 M_\odot$ neutron star, $R_{1.4} \sim 14$ km,
which is disfavored by the constraints from NICER~\cite{Mill19} and GW170817~\cite{Abbo18}.
In contrast, the radii of a $1.4 M_\odot$ neutron star obtained using the TM1e and BigApple
core EOSs are more consistent with current constraints, which are related to their small
slope parameters.
One can see that the mass-radius curves in the TM1e and BigApple cases
go past the maximum mass star configuration, where the maximum masses predicted by
TM1e and BigApple are about $2.12 M_\odot$ and $2.60 M_\odot$, respectively.
It is shown that the maximum neutron-star masses can be significantly affected
by the core EOS at high densities, while the influence of the inner crust EOS is
almost invisible. In all cases, the results are compatible with
the mass measurements of
PSR J1614--2230~\cite{Demo10,Fons16,Arzo18},
PSR J0348+0432~\cite{Anto13}, and
PSR J0740+6620~\cite{Crom19},
but only BigApple core EOS can support a $2.6 M_\odot$ neutron star.

The dimensionless tidal deformability of a neutron star is calculated from
\begin{eqnarray}
\label{eq:td}
\Lambda=\frac{2}{3}k_2 \left(R/M\right)^{5},
\end{eqnarray}
where $k_2$ is the tidal Love number which is computed together with the TOV equation~\cite{Ji19}.
In Fig.~\ref{fig:14LAMB}, we display the dimensionless tidal deformability $\Lambda$
as a function of the neutron-star mass $M$.
The results using the inner crust EOS with $L=40$ and 113 MeV are shown by thick and
thin lines, respectively.
The influence caused by different crust EOS is small,
but the core EOS can significantly alter the tidal deformability $\Lambda$.
The results of $\Lambda$ using the PC-PK1 core EOS are much larger than
the constraints from GW170817~\cite{Abbo18} due to its large symmetry energy slope.
The TM1e and BigApple core EOSs predict relatively small tidal deformabilities
which are more consistent with the constraints from GW170817~\cite{Abbo17,Abbo18,Fatt20}.

\section{Conclusions}
\label{sec:5}

In the present work, we have studied the properties of nuclear pasta phases,
which may occur not only in the inner crust of neutron stars but also in
stellar matter with a relatively large proton fraction.
We have performed fully three-dimensional Thomas-Fermi calculations
in a cubic box with periodic boundary conditions.
The calculations were carried out for both cold stellar matter with a fixed proton
fraction and neutron-star matter in $\beta$ equilibrium.
For the nuclear interaction, we have employed the RMF-PC approach
with the PC-PK1 parametrization, which could provide a good description of
ground-state properties for the nuclei all over the nuclear chart.
In order to examine the influence of nuclear symmetry energy and its slope
parameter $L$, we have generated a set of models with different $L$ at saturation
density based on the PC-PK1 parametrization by introducing an additional coupling
term between the isoscalar-vector and isovector-vector interactions.
All generated models have the same isoscalar properties and fixed symmetry energy
$E_{\rm{sym}}$ at $\rho_B=0.12\, \rm{fm}^{-3}$, which ensure to provide
similar binding energies of finite nuclei with the experimental data,
but they have different density
dependences of $E_{\rm{sym}}$ that may play a crucial role in determining
nonuniform structures in the neutron-rich matter.

We have investigated the pasta structures in nuclear matter with a fixed proton
fraction $Y_p$ by using two limit models with $L=40$ and 113 MeV.
It was found that these two models provide similar features of pasta
structures for $0.3 \leq Y_p \leq 0.5$, but significant differences could
be observed in the low $Y_p$ region, where the symmetry energy is expected to
play an important role. Generally, the ground state of nonuniform matter
at low densities is a bcc lattice of droplets, while nonspherical pasta phases
may appear at the density $\rho_B > 0.02\,\rm{fm}^{-3}$.
The onset density of nonspherical shapes increases as the proton fraction
$Y_p$ decreases. In our three-dimensional calculations, typical geometric
shapes like droplets, rods, slabs, tubes, and bubbles were observed
before the transition to uniform matter. In addition, some intermediate
structures around the shape transition were also observed, which would make
the transition between different shapes more smooth.

For neutron-star matter in $\beta$ equilibrium, we found only spherical droplets
were formed before the transition to uniform matter using the models
with $L=40$ and 113 MeV. The values of $Y_p$ obtained in $\beta$ equilibrium
are very small, which lead to an early onset of uniform matter.
The crust-core transition occurs
at $\rho_B \simeq 0.072\, \rm{fm}^{-3}$ with $L=40$ MeV and
at $\rho_B \simeq 0.057\, \rm{fm}^{-3}$ with $L=113$ MeV.
We studied the correlations between the symmetry energy
slope $L$ and the crust-core transition by employing the set
of generated models. It was seen that both the baryon density and
the proton fraction at the crust-core transition decrease with
increasing $L$. The resulting EOSs of the inner crust using the models
with $L=40$ and 113 MeV were applied to study the properties of neutron stars.
It was shown that massive neutron stars are insensitive to the inner
crust EOS, while visible differences could be observed in the radii of
low-mass neutron stars.
We emphasize that although nonuniform structures in the inner crust
have less influence on the bulk properties of neutron stars,
they may be important for interpreting cooling observations.
In addition, the properties of pasta phases in supernova matter
would affect the neutrino signal, which need further investigation.

\section*{Acknowledgement}

This work was supported in part by the National Natural Science Foundation of
China (Grants No. 11675083 and No. 11775119)


\end{document}